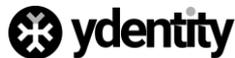

# Decentralized Nation: Solving the Web's Identity Crisis

*"In the digital world, the most valuable things I have are
who I am – my identity – and what I do – my reputation!"*

Frederic A. Jumelle[1], Timothy J. Pagett[2], Ryan S. Lemand[3]

Ydentity Organization[1,2], Neovision Wealth Management[3]

{f.jumelle[1],t.pagett[2]}@ydentity.org   ryan.lemand@neovision-wealth.com[3]

**Abstract**

The Web of today whether you prefer to call it Web 2.0, Web 3.0, Web 5.0 or even the metaverse is at a critical stage of evolution and challenge – largely centered around its **crisis of identity**. Like teenagers who cannot assess properly their reason for being and do not seem ready to take responsibility for their actions, we are constantly blaming the very system we are trying to get away from. To truly realize the benefits from innovation and technology – this crisis has to be resolved – not just through tactical solutions but through developments that enhance the sustainability of the web and its benefits.

Significant strides are being made in the evolution of digital services enabled by technology, regulation, and the sheer pace of societal change. The journey to the decentralized Web is mirroring the **convergence of the physical and digital** worlds across all economies and is increasingly embracing the digital native world. Technology has provided the foundational platform for individuals and entities to create and manage wealth, potentially without the need for big institutions. Ironically, despite all of the advancements, we are still facing an unprecedented and increasing wealth gap. At the core of this, we believe, is the **imbalance of power** – a physical world problem that has now been conveniently replicated in a digital version of that world representing a more pervasive threat to our democratic existence much more so than that we faced within our existing physical world.

Clearly, the system is broken – not just around the edges but at the very core of the democratic underpinning of our society.

In this whitepaper, we propose how artificial intelligence on blockchain can be used to generate a new class of identity through direct human-computer interaction. We demonstrate how this, combined with new perspectives for sustaining community and governance embedded within the use of blockchain technology, will underpin a sustainable solution to protect identity, authorship and privacy at the same time while contributing to **restore trust** amongst members of a future decentralized nation and hence contribute to solving the Web's most significant identity crisis.

## Decentralized Nation:
## Solving the Web's Identity Crisis

From birth, each individual has the right to an identity. The identity of an individual is the assertion of their existence in a society. It is also a matter of recognition of their individuality and what differentiates them from their peers. Having an identity is a fundamental human right which allows each individual the ability to enjoy all of their rights. [1] Exercising and maintaining identity as a **fundamental human right** is a critical challenge in cyberspace balancing the need for a digital version of identity that allows access to technology and the services provided by technology to mediate relationships.

Early generations of digital identity do little more than capture a person's identity dimensions as dictated by the information required to support Know Your Customer expectations (name, address, education, credentials etc.) driven by regulators, marketers and product developers. This so called **Personal Identifiable Information (PII)** has increasingly become the subject of regulation and protection from abusive practices that are particularly prevalent within the web. Our next generation identity technology, Ydentity ID, captures signals directly from the user without the need for capturing so called PII and computes the attributes of this natural person to define and serve them better.

In this paper, we will explain to you why you need to act now and think differently to remain relevant in the future. **A new digital democracy is coming – with identity at the core.** We believe that Ydentity technology gives users the right to redefine civil society [2] in the digital form of a decentralized nation. Through connectivity and empowerment, we can together recreate the original democracy we always wanted to be part of, and the Web will rise again.

## 1. Legal Identity

A critical place to start any consideration of sustainable and true identity in cyberspace is with the basic principles that underpin identity from a legal perspective.

### 1.1 Identity of a natural person

This right to an identity is enshrined in Article 6 of the Universal Declaration and reiterated by QHRC Section 15 of the Human Rights Act 2019:

**"Everyone has the right to recognition everywhere as a person before the law."** [3]

From birth, each individual has the **right to an identity**. The identity of an individual is the assertion of his or her existence in a society. It is also a matter of recognition of their individuality and what differentiates them from their peers.

Having an identity is a fundamental human right which allows each individual the ability to enjoy all of their rights.

Identity encompasses the family name, the surname, date of birth, gender and nationality of the individual. Through these details, an individual will hold rights and obligations specific to their status (woman, man, child, handicapped, refugee, etc.).

From birth, each individual has the right to have a name and a surname. In most countries parents have a duty to declare the name, the surname and date of birth of a newborn to the authorities in charge. By recording this birth, the State officially recognizes the existence of the child and formalizes their status in the eyes of the law. In addition, through this name and recording in the Registration of Births and Deaths, the child will be able to establish filiations, that it is to say links of blood relations linking the child to their father and/or mother.

From birth, the child also has the **right to a nationality**. Nationality can be obtained in two different ways:

- *Jus sanguinis (by blood)*: the child will have the same nationality as his/her parents.

- *Jus soli (by birth)*: The child will have the nationality of the territory on which he/she was born, even if his/her parents have a different nationality.

Nationality is confirmed through the issuing of a **birth certificate**. It is an important aspect of a person's life, as it is an attribute of citizenship. Nationality allows establishment of the affiliation for an individual to a nation.

## 1.2 Management of individuals' identities

Separate from an individual's right to an identity is the concept of state authority over its citizens. Every country has some form of registering an individual as part of its management and control over society whether through birth certificates, passports, ID cards, national insurance numbers or any combination of the above.

This confirms an **individual's status as a citizen** of the relevant jurisdiction but also serves the societal function of allowing a state to track its citizens' activities through their actions and financial transactions. The tension between the individual's right to freedom and privacy and the state's reasonable requirement to monitor its citizens for security, protection and other purposes is at the heart of one of the most important philosophical debates of our time.

Currently, more than half of the world's countries operate a National Identity Card System. Some countries use biometric ID cards, others have biometric passports. Currently India has one of the most advanced and comprehensive systems of national ID called Aadhaar. [4]

## 1.3 Refugee/migrant problem

World Bank ID4D Global Dataset 2017 suggests that 1.1 billion people worldwide do not possess sufficient documentation or transactional evidence to prove their own identity to a level sufficient to be acceptable for regulatory standards.

This gives rise to economic and social consequences for the person lacking a formal identity as formal identification is usually required to access healthcare, education and financial institutions as well as for gender equality. Similarly, a proper identity register is necessary for governments to provide social welfare services. It is widely considered to be one of the greatest contributors to continued financial exclusion.

In 2015, the UN adopted the 2030 Sustainable Development Goals and **Goal 16.9** [5] sets the provision that by that year the UN shall provide legal identity for all persons including free birth registrations.

This is widely recognized as a challenging commitment with debate continuing at pace across and between many signatories as to what constitutes legal identity. There is some hope that digitization may provide the basis for accelerating the achievement of this objective – in the form of a hybrid sovereign identity solution. However, skepticism remains as to the true intention of a universally applied identity – be it digital, hybrid or physical.

## 1.4 Digital identity

A critical challenge in cyberspace is knowing with whom one is interacting. Using static identifiers such as passwords and email there is no way to precisely determine the identity of a person in a digital space, largely because this information can be stolen, manipulated or used by many individuals acting in concert or as one. To overcome this, digital identity has evolved based on dynamic entity relationships that can be captured from behavioral history across multiple websites and mobile apps can verify and authenticate an identity with high accuracy.

By comparing a set of entity relationships between a new event (e.g., login) and past events, a pattern of convergence can verify or authenticate the identity as legitimate where divergence indicates an attempt to mask an identity. Data used for this type of digital identity is generally anonymized using a one-way hash, thereby avoiding privacy concerns. Because it is based on behavioral history, a digital identity is nearly impossible to fake or steal.

A digital identity is information on an entity used by computer systems to represent an external agent. That agent may be a person, organization, application, or device. ISO/IEC 24760-1 defines identity as "set of attributes related to an entity". [6]

The information contained in a digital identity allows for assessment and authentication of a user interacting with a business or social ecosystem on the web, without the involvement of human operators. Digital identities allow our access to technology and the automated services they provide and make it possible for technology to mediate all forms of personal and business relationships.

The term **"digital identity"** has also come to denote aspects of civil and personal identity that have resulted from the widespread use of identity information to represent people and their actions in digital form. Digital identity is now often used in ways that require data about persons stored in digital form to be linked to their civil, or national, identities. Furthermore, the use of digital identities is now so widespread that many discussions refer to *"digital identity"* as the entire collection of information generated by a person or entity's online activity.

This includes **usernames and passwords**, online search activities, birth date, social security, and purchasing history. Especially when that information is publicly available and not anonymized and can be used by others to discover that person's civil identity.

In this wider sense, a digital identity is a version, or facet, of a person's social identity. This may also be referred to as an online identity.

The legal and social effects of digital identity are complex and challenging. However, they are simply a consequence of the increasing use of technology, and the need to provide technology with information that can be used to identify external agents.

ID2020 [7] considers the core features and advantages of digital ID to be as follows:

- personal: unique to the owner and the owner only.
- persistent: lives with the owner from birth to death.
- private: only the owner can control his/her own identity, and the owner can choose what to share and with whom.
- portable: accessible anywhere the owner happens to be through multiple form-factors.

Digital identity fundamentally requires digital **identifiers – strings or tokens** that are unique within a given scope (globally or locally within a specific domain, community, directory, application, etc.).

Identifiers are the key used by the parties to an identification relationship to agree on the entity being represented. Identifiers may be classified as omnidirectional and unidirectional.

Omnidirectional identifiers are intended to be public and easily discoverable, while unidirectional identifiers are intended to be private and used only in the context of a specific identity relationship.

Identifiers may also be classified as resolvable or non-resolvable. Resolvable identifiers, such as a domain name or email address, may be dereferenced into the entity they represent, or some current state data providing relevant attributes of that entity.

Non-resolvable identifiers, such as a person's real-world name, or a subject or topic name, can be compared for equivalence but are not otherwise machine understandable.

There are many different schemes and formats for **digital identifiers**. The most widely used is Uniform Resource Identifier (URI) and its internationalized version Internationalized Resource Identifier (IRI), the standard for identifiers on the World Wide Web.

Lightweight Identity (LID) and OpenID are web authentication protocols that use standard HTTP URIs (often called URLs), for example. [8] [9]

### 1.5 Privacy and security

Businesses and governments, alike, fight for control over access to **user information** in order to maximize advertising dollars, maintain political power, optimize their positioning within, and sometimes even monopolize, marketplaces. Until the rise of the sharing economy platforms, large traditional businesses appeared to maintain a complete monopoly in a variety of industries, such as tourism, hospitality, retail and personal banking, investing and capital markets, and taxi/ride hailing services.

Following the revelations of Edward Snowden on mass surveillance by the NSA [10] in 2013 and the Facebook-Cambridge Analytica data collection scandal [11] in 2018, many businesses have, through a combination of peer, regulatory and social pressure, shifted **power back toward the consumer** through allowing individuals to assume the role of these entities in peer-to-peer transactions. Already, the sharing economy has seen extremely rapid growth and will continue to become a part of everyday life.

Despite the disruption the sharing economy has generated, there is still significant control over wide-spread user information in the hands of much fewer private businesses. While the sharing economy has helped to reduce the power of large, traditional businesses, individuals must still place a lot of trust in the security promised by the organizations who design, create and build any given sharing platform. New technologies continue to emerge, however, affording more and more individuals even greater and more secure control over their own data and digital presence.

### 1.6 Reputation services and the blockchain

Usually, user ratings and reviews are tied to a single platform which owns their content and individuals have to start from scratch if they want to build their reputation on another sharing platform. Reputation response can be set for email identifiers and used by a reputation services provider. [12] Thus, third party platforms afford users greater control over their data by allowing them to utilize their entire digital reputation as leverage to prove their trustworthiness across peer-to-peer marketplaces.

Users can voluntarily connect their social media and sharing platform accounts in order to use their trust profile which is based on the ratings and reviews they have received on connected accounts. All of this happens when users sign up to a new platform and third-party technology extends the sign-up flow with a seamless integration where users can build their trust profile.

Mixing **user identity and rating services** with blockchain technology could result in an even better final product.

As mentioned above, ratings and reviews are usually related to a specific service and are owned by the company/person behind it. The structure of the current digital reality is based on clusters of information, made available by large companies.

**Online existence has historically been dependent on the access provided by centralized services**, and our resulting digital identities and reputation are still owned by a handful of organizations that provide these services.

We pay them with information about ourselves, oblivious to how that information is actually used (and by whom).

It is probably fair to say that the organizations that have truly monetized the value of data held in digital form are the BigTech businesses that control access to technology.

However, in a move that could potentially challenge the business model of data ownership, blockchain (or distributed ledger) technology provides an alternative way of storing information without the need of a central authority whilst still serving and connecting users.

As an example, the Bitcoin blockchain stores information about the balance of each Bitcoin address.

The same approach could be applied to many other types of information. Data about anything that could be represented digitally can be stored on a blockchain and made accessible by anyone with the proper decipher tool. Immutable, public and, at the same time, encrypted information could be the key to enhancing identity and reputation solutions.

Solutions that provide users with the ability to have improved control over their digital privacy and the ability to sharing that personal information only when they choose to or are legally required to are a critical need as Web 3.0 evolves.

### 1.7 Ydentity, a decentralized ID

Traditional forms of digital identity generally do little more than capture the necessary components required to support compliance with regulatory requirements, to support models and analytics driven by digital platforms and ecosystems or to aid in social connectivity. Most, if not all comprise digital representations of physical documentation (e.g. name, address, education, credentials).

Ydentity's creator, Dr Frederic Jumelle, considers that this does not capture the attributes of a natural person therefore does not define nor serve the person anywhere near as well as it serves the entity requiring the identity.

Ydentity uses a method and system for neuropsychological performance testing based on a **self-interview on portable device** during which device sensors capture the user's cognitive and emotional responses to a selection of 30 questions selected randomly from a proprietary database of over 600 questions. The Artificial Intelligence (AI) comprises a cellular architecture with several artificial neural networks for **processing the signals** captured by sensors during the interview. Ydentity uses an edge computing approach which includes the preselection and preprocessing of most informative data directly on the device and the destruction of all materials after processing. AI modules have been pre-trained using international human emotions datasets and will benefit of online learning when new profiles are generated.

Each user's metadata file comprises **7 scores** (tolerance, credibility, maturity, autonomy, emotional state, worthiness, and w-range in reference to the tested population) along with 5 demographics (age, gender, ambition, job level and education level) and their YDR reputation level.

A record of the user's **128-dim matrix of facial features** is also uploaded onchain to allow an automatic matchmaking between the record and the user's face at login preventing duplicate and deep fakes. Each Ydentity ID can be used to check identity and authenticate, authorize, certify and upload/download documents on IPFS, or create a digital twin for risk management and protect the rights of the user/owner across various blockchains.

The Ydentity proprietary login system includes **2 layers of anti-spoofing technology** based on video capture using Multi-task Cascaded CNN for gatekeeping unique access to the dApp and wallet of the user, making it a Self-Sovereign and Decentralized Identity. [13]

### 1.8 The Ydentity.org Association

The Ydentity.org Association was created in Switzerland, Canton de Genève, [14] to guarantee and protect the right of every natural person or citizen to obtain, own and use a Ydentity non-fungible token ID.

While the Ydentity dApp enables users to self-generate an individual Ydentity based on signal capture and processing of the attributes of each user by an artificial intelligence unit, the association is lobbying to:

- facilitate the use of such an Alt-ID by every citizen for identification and authentication during internet navigation; and

- create and maintain a database of active members that can represent the world population in terms of diversity to support the development and generalization of Alt-ID in the fields of online identification of persons, KYC, and risk management oracle services; and
- warrant the principle of neutrality and independence related to individual identity.

The Association is an **association of stakeholders and followers** and has no profit purpose.

The Association may pursue all lawful activities to achieve its purpose. In particular, the Association may undertake the active and international recruitment of its members to meet its purpose; advertisement and development of a marketing strategy for its purpose; the formation of a **lobby to influence** governments, non-profit organizations and other associations in relation to its purpose and implementation.

### 1.9 The right to own a decentralized ID

A Ydentity token ID is a non-fungible-token that can be **generated on decentralized applications** deployed on the Ethereum and Rootstock blockchains. Owners are able to use it to identify, communicate and interact with one another.

A Ydentity **token ID** can be used to create a unique decentralized Identity Document that is not transferable and cannot be traded. It contains a PII-free file (without Personally Identifiable Information) that is uploaded on-chain when minted. This on-chain file is non-encrypted and can be searched on the dApp by other Ydentity users. The result of a search is a list of token addresses that can be used to connect via secret email (ex: ethmail.cc) and build a Ycommunity. It can also be used by third parties for other purposes such as statistics, polls, decision-making protocols, etc.

Each Ydentity token is held in its owner's wallet along with Ydentity reputation tokens (YDR).

If a Ydentity ID owner decides to burn his/her Ydentity ID, there will be consequences resulting from this action up to being removed from the Ycommunity or the DAO in certain cases. Reinstatement would be possible but subject to a new full registration and Network Governance approval.

### 1.10 YDR, the new currency of trust

If identity is **who I am** then reputation is **what I do** – with my identity.

Each Ydentity ID owner can earn reputation in the form of Ydentity reputation tokens (YDR). [15] These ERC20 tokens are earned when owners share their unique Ydentity signature (proof-of-identity) for example when underwriting transactions. The number of **YDR tokens** awarded for a transaction will depend on the smart contract underpinning the transaction especially the definition of the criteria of satisfaction between parties.

Ydentity ID owners will find ways to spend their YDR in the DAO for example while searching for peers and building Ycommunities or joining certain focus groups, clubs or buying privileges to invest in complex products or borrowing cryptos or gaining access to certain social graphs.

A certain level of YDR tokens will be necessary to join some groups but also multiply the gain from these groups. In the case where a Ydentity ID owner behaves poorly, YDR tokens can be lost or cancelled, and the reputation level of this individual will decrease. Because of this loss of status, the individual can be denied access to certain groups or services until they regain tokens. Reinstatement will be decided by Network Governance especially in cases where the reputation level has reached zero.

## 2. Ydentity Technology

### 2.1 Ydentity (NFT) token ID

Ydentity is a non-fungible token identity that stores the person/owner's attributes on a digital ledger or blockchain which certifies that this digital asset, the user's Ydentity, is a unique **decentralized identity** (DID) [16] and a **self-sovereign identity** (SSI). [17][18]

Each Ydentity token ID is by definition unique, non-fungible, ownable, non-transferable and burnable. It tokenizes the physical, cognitive and affective attributes of a person captured from/directly on the user by himself/herself during a self-administered and timed Video Emotion Recognition-based Interview or VERI.

The Ydentity DAO stage will be a platform offering members the chance to mint their Ydentity IDs and start interacting with other holders, building groups, clusters and **Ycommunities**, to earn and spend YDR in numerous dApps deployed by the Ydentity Foundation and other partners of the Foundation licensing the Ydentity technology. The DAO can also invite developers to build dApps for the DAO, improve its algorithms, accept challenges for which they receive grants in the form of YDR tokens.

### 2.2 Ydentity in a decentralized economy

Ydentity ID will become the key element of risk and decision management of a decentralized economy. **Risk and decision management** are important parts of both an individual's and a corporation's daily existence. We all practice risk and decision management to varying degrees and for corporations' assessment and mitigation of risk has been performed for decades. Entities, whether individuals or corporations, have been classified in broad categories regarding their relation to risk. Risk managers call upon these categories throughout their career when assessing risk, mitigating risk, selling products and the acceptance of the risk in the first place.

Introducing a new type of attribute-based digital identity with a stack of scores available for virtual interactions is a must that could outsmart the system of physical and digital identity cards. Similarly to a government-issued ID, all kind of rights can be attached to a Ydentity ID but conversely Ydentity can keep a log of background history and offer legacy rights to the owner(s). Developing the technology where **users can build their own attribute-based identity**, in the form of an immutable digital file, own it to earn and trade while sharing only a unique signature was a necessary step to protect the uniqueness of one's identity.

An NFT is a non-fungible unit of data stored on a blockchain that can offer a representation of the user during web sessions. Users can earn reputation tokens (fungible) during transactions since trust is removed from the equation, only reputation matters.

The Ydentity profiling framework using a **Video Emotion Recognition Interview** [19][20] was developed in partnership with a research team of the Hong Kong University of Science and Technology – Neuromorphic Interactive System Lab, and is constantly upgraded to allow users to self-assess and become their own oracle to predict the effect of their (potential) decisions on communities; or evaluate the systemic effect of their decisions on platforms, graphs or markets; or on pricing of products they want to trade in before taking the decisions; or evaluate co-investing risks and the implication of sharing risks on certain products or services before, during and after taking decisions.

Blockchain **oracle services** currently include many prominent projects and numerous decentralized finance (DeFi) dApps that could work with Ydentity API and benefit from the Ydentity technology at large while using onchain data as first party oracle for risk management services.

### 2.3 Neuropsychological testing at the core

The technology contributing to the generation of Ydentity ID is a patented method and system for neuropsychological performance test. [21][22][23][24] It is based on a terminal device used to interact with a cloud server which only stores user processed data and is logged into by the users through their terminal device. A test module comprises the user information, which is stored in the cloud server or can be downloaded from the cloud server and is directly accessed through said terminal device and is trained by the artificial neural network. User information comprises user performance metrics and the terminal device can display the neuropsychological performance test results before they are minted to a non-fungible token ID.

Ydentity technology can be applied to pre-screening, remote screening and onboarding of human resources; sorting of personal accounts, fraud prevention and forensics for social media; matchmaking in client relation management and dating; onboarding and remote onboarding of new customers in the financial services industry for compliance Know Your Client ( "KYC" ) or Customer Due Diligence ( "CDD" ) regulations and also any new virtual services to persons including providing smart ID for smart cities. Compatible with other identification and identity authentication/verification technologies, Ydentity technology allows the **creation of true personal identity by using personal metrics with a high degree of accuracy and security**. Ydentity technology enables some of the biggest challenges of the internet such as the large amount of fake accounts including the fake social media accounts that can pose a threat to society to be easily solved.

**2.4  Recognizers on edge to enhance privacy**

Based on a grant for pioneer research titled "Bandwidth Aware Video Emotion Recognition in the Cloud", we are developing **new recognizers** for the Ydentity dApp frontend that can enable edge computing of the captured signals and guarantee privacy in 5G or 6G environments.

Our arousal network (shallow neural network) can pick up the key frames in the interview package and deliver the total emotion metric. It is more effective to provide measurements of emotional states that are more finely grained than discrete categories. It also works from video input.

The team at Hong Kong University of Science and Technology has an established track record in applying deep neural network (DNN) technology to address this area (Deng et al., 2020; Deng, Chen & Shi, 2020; Zhou, Pi & Shi, 2017; Zhou & Shi, 2017). [25][26]

Deep neural networks developed by the team have yielded state of the art performance on international benchmarks, e.g. winning international challenges, such as the 2017 Facial Expression Recognition and Analysis (FERA) AU Intensity Estimation Challenge (Zhou, Pi & Shi, 2017) and the **Valence-Arousal** and the Action Unit Tasks in the 2020 Affect Recognition in-the-wild Challenge (Deng, Chen &Shi, 2020). [27] Optimization also includes Iterative Distillation for Better Uncertainty Estimates in Multitask Emotion Recognition. [28]

In particular, the team has developed technology that classifies emotions along the continuously valued dimensions of valence and arousal, which can achieve any value between -1 and 1, rather than only seven discrete values (Deng et al., 2020). This allows for the more fine-grained distinctions required by user profiling.

It has also developed algorithms for **Facial Action Unit** intensity estimation (Zhou, Pi & Shi, 2017). Action units are localized facial muscle movements, which can be used to evaluate and describe human mental states like depression and happiness. The problem of intensity estimation seeks to determine the extent to which a particular action unit is activated, as a number ranging anywhere from zero to one. This fine-grained estimation provides more information than just a single binary detection result of whether or not the action unit is present.

**2.5  From third party oracle to first party oracle for third parties**

Much of the evolution of traditional finance has increasingly involved the development of data rich experience driven models – in many cases provided through the use of third-party oracles. Take the provision of lending finance driven by the scores generated and held by credit reference agencies, where propensity to repay is inferred by statistical correlations between actions, observable and collected attributes and outcomes. The same is true in the behavioral and statistical models that underpin much of the algorithmic trading that dominates many global investment and trading markets.

In both of these cases risk management is based on statistics, a stochastic approach using a limited number of variables that are captured from markets by sensors and feed agents which apply algorithmic relationships between input and output.

AI/Machine Learning is being used with varied success to overcome limitations in statistical using big data (de-personalized and anonymized data) in sufficiently large volumes, with high velocity and variety. This brings the benefit of correlations and the power to take decisions with confidence even though the accuracy at an individual node level is generally considered mediocre. This accuracy challenge can be addressed through AI/ML using personalized data in smaller volumes, but high accuracy would prove very impactful. However, this would contradict much of the privacy protection regulation and therefore cannot be used or is just too expensive to overcome.

In decentralized finance or DeFi, many of the apps that have been developed are looking to serve the users on a peer-to-peer basis without the need to challenge privacy regulations and to avoid the cost of relying on third party anonymized data such as credit reports. Notwithstanding the way these models are configured they still rely heavily on historical observations.

We are seeing emerging challenges with the DeFi apps and the inherent limitations of basing behavioral models on static points in time actions and attributes rather than dynamic analysis combined with emotional positioning and response. The latter are just too expensive to capture and analyze.

Despite the existence of Nobel winning research from the likes of **Daniel Kahneman and Richard H. Thaler** that provides definitive support that emotions play a critical role in driving decision making, particularly in finance, many of the models in use in finance applications today, whether traditional, digital or DeFi, do not consider emotions as an input.

The Ydentity token ID incorporates the emotional component with the cognitive and timing components of decision making to provide a unique combination of attributes for each Ydentity holder. When held in sufficient volumes the population of Ydentity holders can form the basis for assisting them to **"self-appraise their relative risk level"** whilst at the same time as empowering them to share this for their own benefit such as getting a better loan rate or accessing a selection of equity funds.

The technology and methodology supporting the Ydentity token ID, is an effective "first party" input to allow "third party" indefinite real time risk profiling as well as allowing users to decide whether to link users' data to the market and get feedback for online learning without interruption. Ydentity can be used to provide risk management data and first party oracle services in peer-to-peer interaction and applications without the need for costly intermediaries leading to enhanced profit opportunities.

**2.6  Ydentity onchain data for DeFi (by third party dApps in the DAO)**

Ydentity is based on a technology using AI neural networks for profiling and risk and decision management. In this way, it can be used as a critical input to a model contributes to **predict customer's propensity to uptake a product and predict the correlated risk of certain products**. Through the unique and immutable features of Ydentity, we believe that it could be used to set up a feedback mechanism to learn directly from the users. From a reputation point of view, the actions of the user could be used as a trigger alert for management to take timely mitigation actions.

Although Ydentity can currently be used to provide highly accurate predictions based on neural networks, the method and system (especially the hidden layers or LSTM mechanisms) remain difficult to explain to the majority of regulators therefore it cannot currently be used to achieve regulatory compliance and for reporting to regulators. However, we believe that this will change as the use of this type of approach grows exponentially on the Web.

Ydentity has the potential to be used for marketing and decision making such as business decisions with confidence without having the need for total accuracy. Because companies' profile, user's risk profiles and markets can evolve, **recalibration with online training** of the engines is important in risk management and is Ydentity's main feature. AI/ML on personal data enables first party oracle services provided directly to the user from the users' data, standalone or for pooling. Ydentity can contribute data and be a first party oracle on its own regarding pricing, underwriting and modelling individual or group decision. It can possibly target systemic impact and analytics at an advanced stage.

### 2.7 First party oracle and sustainable value creation and transfer

Human decision making applies the decision theory while human planning applies the **theory of choice** and **game theory**.

A first party oracle is fed by the user(s) for self-serving interest and authorship such as optimization of decision making, planning and value transfer based on:

- **Interaction** between decision makers, i.e. co-investing Ydentity owners.
- **Intertemporal choice**, the Ydentity owners enter the Web3 sphere at different times.
- **Non-zero-sum game** theory applied to Web-based investment markets where the sum is not zero because the interaction between decision makers is generating a new intrinsic value comparable to the enthalpy of formation which can take the form of a reward in YDR reputation tokens and other earnings and enable value transfer. In the opposite direction, penalties can be given in a form related to the enthalpy of dissolution. This feature will be developed as a test against a zero-sum game also called a Ponzi Scheme.

Any mechanism for sustainably creating value from one's identity needs to simultaneously consider the mechanism for the ultimate transfer and realization of that value.

The value of identity in almost all existing Web 2.0 applications is not based on what you do with your identity – it is more based on what you did with your identity – and the attributes that "define" your identity. You are considered more valuable if you did the things that models/AI said were valuable – borrowed money, paid it back, bought certain goods and services, accumulated wealth by a certain age, lived in certain areas, had certain ethnicity, went to the right schools, got the right job. All attributes that are dictated and measured from the "center" – controlled by those that have sufficient data to define what is "right".

However, consider a **digital native world** where what you do, where you go and how you act are only capable of capture in a digital form and are really only controlled by what you want to do. In both this and the converging digital and physical world, digital currencies, crypto currencies, tokenization and fractional ownership are being progressively recognized as mechanisms for creating both incentive and reward – providing a new basis for creating "transferable" or "realizable" value without the need for traditional monetization.

In this world – what is considered good?

Much of the progressive thinking on this tends toward the creation of smart contracts that recognize and reward "good" behavior through granting access of digital rights/assets whilst simultaneously penalizing bad behavior – largely through loss of access or digital rights/assets. Through empowered users, whose positive choices become the self-fulfilling governance mechanism to weed out and isolate bad actors, the need for centralized control becomes obsolete.

Reputation – what you do – becomes equally as valuable in a digital world as who you are.

Add a mechanism to transfer the **value of reputation** to create even more value – only by successful contractual actions – one can never "buy" reputation but only earn, use and lose reputation. The velocity of usage will provide the momentum for new ways of working, where you are only rewarded for the contribution that you make – no entitled wealth creation – just reward for performance considered through consensus to be of value to the community. This has the basis to fundamentally change the way in which reward and incentive mechanisms operate, not just in the digital native world but in the physical world as well.

This is analogous to the current positioning that **copyright and intellectual property** plays in both the physical and digital world. The recognition of effort – both tangible and intangible – is given credence through the successful application of "protection" – allowing for the owner of that protection to not only capture appreciation of the intangible value of the reputation being protected, but also through royalties paid for the use of that intangible value by others. Equate this to a digital native world and effectively your reputation is the intangible value of how you act in the digital world. However, in a digital native world there is no "ready-made" mechanism to protect the value of your reputation or what you do with your reputation – unless it is attached to an immutable form of identity that is protected and controlled by the owner of that identity. A logical extension of this is then the real future of work, where contribution to work is proven by the consensus mechanisms inherent within blockchain technology. When combined with immutable identity there is a new way of recognizing and rewarding contribution – creating tangible rewards and intangible value – in the same way, but far more efficiently, than was created in a physical world. The value from what you do, not from who you are.

What you do has no value if it is not backed by the **authenticity of who you are**. It is critical that identity be simultaneously verifiable and recognizable. Value transfer mechanisms need to be based on a better version of identity – one that represents the convergence of the physical and digital worlds – through human computer interaction. Not one developed on a static basis, that is really just a digital representation of a physical form but one that is a dynamic representation of the core combination of an individual's identity – the physical, cognitive and emotional elements working in concert. Not one that is controlled centrally, but one created and controlled by the true holder of that identity – on an immutable basis – for the good of only that holder.

## 3. Decentralized Nation as a Solution

### 3.1 Definition of a nation

A nation is a large body of people united by common descent, history, culture, or language, inhabiting a particular country or territory, according to Oxford Languages and Google.

In other dictionaries, **nation** refers to:

- All human beings living in the same territory, having a community of origin, history, culture, tradition, sometimes language, and constituting a political community; or
- Abstract, collective and indivisible entity, distinct from the individuals who comprise it and the holder of sovereignty; or
- In biblical literature, "nations refer to pagan peoples, as opposed to the chosen people."

### 3.2 Definition of a decentralized autonomous nation (DAN)

A decentralized nation is likely to be an **abstract, autonomous, collective and indivisible** organization distinct from the individuals who comprise it and who are the holders of sovereignty.

As a new type of nation pioneering in a field of abstract concepts, a decentralized nation should look for the best available systems of political, social and economic governance before inception in order to avoid the pitfalls and failures of systems that are proven inefficient, oppressive and misleading. We believe there is a sufficient number of indicators for such failures to identify what works and what is not working. The interest of looking for mathematical formulation of systems is that mathematics allow computation and computation can automate the rules of execution of contracts.

### 3.3 Why do we need one or more DANs?

Let's start with the definition of **"Security"**:

- a state of being free from danger or threat; or
- something deposited or pledged as a guarantee of the fulfilment of an undertaking or the repayment of a loan, to be forfeited in case of default.

Are the two definitions equivalent or ambivalent or something from the past?

Are all financial assets considered "securities" for the sake of the security of the state which aims to protect its citizens?

A security (definition 2 above) is considered security (definition 1 above) because it is guaranteed by a hard asset, a business that generates or purports to generate cashflow or value, or anything that can be sold to recover a certain amount of money in case of the default of the aforementioned security's issuer, hence where the name came from. In Finance, we use the word "security" to generically describe financial instruments such as stocks, bonds, units in mutual funds, and others. However, in recent years, this definition has become blurred with the arrival of digital assets, namely digital coins and cryptocurrencies.

Many have even argued that cryptocurrencies cannot even be called "assets". In fact, until the arrival of digital assets, in Finance, the words financial assets and securities were frequently used interchangeably.

Many others argue that cryptocurrencies cannot be considered as securities because they fail the Howey test although the Securities and Exchange Commission (SEC) of the United States position seems to be that many cryptocurrencies may be securities based on recent actions they have taken and cases they have brought.

Others have argued that cryptocurrencies are not only assets and securities and more but most importantly a **new way of life**.

In fact, for years now, proponents of cryptocurrencies, have preached that they represent a form of deliverance from political and financial oppression as well as **the end of the surveillance** of our very life by a paranoid "Big Brother".

Cryptocurrencies have become a way out of sorts, a means of escape if you will, that represents the core beliefs of its users giving the opportunity to become their identity throughout and beyond government identification.

Everything that surrounds us and is related to us from a socio-economic perspective has become increasingly represented in a digital form: our medical records in the cloud are readily available to medical practitioners, our credit scores can be consulted by creditors to verify our financial worthiness, our biometric passports contain our biometric data, etc.

However, the usage of fiat currency remains flexible and anonymous if one so chooses. For example, when we use a $20 bill to buy groceries, this transaction will not appear anywhere in the digital world. Hence one of the reasons for developing loyalty cards and other point of sale survey questions such as postcode, phone or name, as a means of collecting information on transactions and customers. You do not have to provide the information – to conclude the transaction if it involves cash. Contrary to this, if the same transaction is made though digital means, contactless smartphone to terminal or a card whether by fiat currencies or cryptocurrencies, it will be recorded somewhere in the digital world and could be traced to a credit card, a wallet, a bank account and to you the owner or holder.

The paradigm of a digitized nation has already been adopted and standardized in many developed countries, as well as in some emerging countries engaged in a rapid digital transformation linked to the digital identity of their citizens.

These digitized nations are centralized around the core aim to connect their citizens' digital identities and data with government services, medical services, financial institutions, and economic policy in a way that can make them proactive rather than reactive. Taken one step further, a **state-controlled digital nation** can decide to assign a social score to each digital identity holder and pretend that is the mirror of the person's actions, behaviors, risk management and financial management. If this social score also includes individual medical records and consumption habits of the person, this person will witness the disappearance of their personality and privacy entirely, the depth of which has only been seen in the most somber hours of humanity.

We can safely say that the industrial revolution has given way to the digital revolution, which started around the year 2000 with the Internet maturing and becoming the backbone of our society and it is still ongoing like a maelstrom swallowing everything known to us from personal to societal, financial, economic, etc.

Where does this leave us?

What will happen to our privacy and individual freedoms if we continue to trade our socioeconomic identity for the sake of spurious security?

Is this first version of digital nation simply creating a heard of sheep that can be steered electronically through push technology?

At Ydentity, we believe that humans need autonomy and freedom to mature and be creative, to have independence and the stamina to make the smart and the right decisions.

We think that **a truly digital nation cannot be deployed by the current governments digitizing their control but by their citizens**, not only the citizens of one nation but the citizens of multiple nations together, transcending the very concept of a nation that has failed so many times throughout history.

There is a possible way between the paradise of the dreamers and the hell of a Kafkian society.

It implies laws made for this new generation of DAN wrapped in legal entities with frameworks designed to guarantee their existence in the middle of the other moving parts of the coming new world order.

It is inevitable because these **DAN represent hope**, the aspiration to have a **better future**, to bring the borders down, to **avoid war**, to save the planet when others are already thinking to leave Earth to Mars or wherever they believe they can have a future without the burden of taking care of we have now in our hands.

### 3.4 Political system and consensus by Proof-of-Authority

A DAN should make the choice of a political system based on the weighted arithmetic mean W which allows the users with higher reputation level to contribute more than the ones with lower level and exclude religious beliefs and leader centricity from the system governance. [29][30]

$$W = \frac{\sum_{i=1}^{n} w_i X_i}{\sum_{i=1}^{n} w_i}$$

where $W$ is the weighted average; $n$ is the number of terms to be averaged; $w_i$ are the weights applied to x values; $X_i$ is the data values to be averaged. A political system allowing weighted **vote** will avoid irrationality in the decision-making process that can integrate with the economy for the purpose of sharing the common good. [31]

With a DAN operated by an organic Ydentity blockchain where trust is distributed, Proof-of-Authority (PoA) for **consensus** [32] will be adopted by the Ydentity network.

In the Ydentity PoA-based network, transactions and blocks are validated by approved accounts, known as validators. To become a **Ydentity validator**, one must own more than 1,000,000 YDR tokens. Ydentity validators run software allowing them to put transactions in blocks. The process is automated and does not require validators to be constantly monitoring their computers, but it does require maintaining the computer ('authority' node) uncompromised. Because individuals earn the right to become validators, there is an incentive to retain the position that they have gained. By attaching good reputation represented by YDR reputation tokens to their Ydentity ID, validators are incentivized to uphold the transaction process, as they do not wish to have their Ydentity attached to a negative reputation represented by a low level of YDR or none.

In simple words, if validators uphold transactions they earn YDR tokens and when they do not they lose YDR tokens.

PoA consensus is considered more robust than Proof-of-Stake because PoA only allows non-consecutive block approval from any one validator, meaning that the risk of serious damage is centralized to the authority node. Robustness of PoA is increased with the number of validators and our recommendation is to have a minimum of 23 validators.

Nevertheless, the results of the most advanced research on the future of blockchain are leaning toward **removing the consensus** mechanism entirely when autonomous entities will be "living" in the network, they will be always right by definition.

### 3.5 Econodynamics

Econodynamics is the application of the mathematics of dynamic statistical mechanics and chaos to the study of economics. [33][34][35] If a DAN is looking for a controlled chaos, it should make a choice of an economic system following the logic of the thermodynamics applied to the economy and be based on **non-zero-sum** game theory. This theory describes situations where one decision maker's gain (or loss) does not necessarily result in the other decision maker's loss (or gain). In other words, it allows creation of a value based on the energy generated by the interaction or transfer of data between parties. The enthalpy is a property of a thermodynamic system that could fit this purpose. An enthalpy change describes the change in enthalpy observed in the constituents of a thermodynamic system when undergoing a transformation. This process is specified solely by their initial and final states.

#### 3.5.1 Enthalpy of formation, energy of creation of bonds also called "connections" in a nation

A common enthalpy change is the enthalpy of formation $H$ or heat of formation which is the change of enthalpy happening when one **substance is formed** from its initial constituents which also increases the entropy of the surroundings.

$$\Delta_r H^\theta = \sum v \Delta_f H^\theta (\text{products}) - \sum v \Delta_f H^\theta (\text{reactants})$$

where $\Delta_r H^\theta$ is the standard enthalpy of formation for a chemical reaction at standard temperature and pressure (STP); $v$ is the coefficient of each respective reactant or product in the balanced chemical reaction; $\sum v \Delta_f H^\theta (\text{products})$ is the sum of enthalpy of each individual product in the balanced chemical reaction; $\sum v \Delta_f H^\theta (\text{reactants})$ is the sum of the enthalpy of each individual reactant in the balanced chemical reaction.

This property could work well to describe the change of energy happening in a non-zero-sum game if the decentralized nation economic system behaves hypothetically like a fully connected system created from the initial constituents of the nation.

#### 3.5.2 Enthalpy of atomization, energy of dissolution of bonds in a nation

Atomization is the change in enthalpy when all bonds between atoms of a compound are broken in a way that they become atoms and are **incapable to recreate the broken bonds**. For diatomic compounds, enthalpy of atomization is equal to enthalpy of total dissociation. This is usually what is happening when a group or an entire society or a nation becomes incapable to maintain the bonds between its members.

#### 3.5.3 Entropy of the nation

Entropy is the **measure of disorderliness** of the system. Entropy generally means disorderliness which is the analog to the variance in arrangements of particles or assets or users. Entropy is represented by S. In case of solid blocks, the users are very close to each other because they are arranged in regular order, so solid blocks have less entropy than liquids. The case of liquid matters or liquidity is the intermediate state between solids and gases in which case users are the farthest away from each other. The more away from each other, the larger and more positive is the entropy. A very organized system has a low or even negative entropy that means there a potential to expand and there are also risks in this expansion if it is not controlled.

$S$(solid/real estate/gold) < $S$(liquid/stock/fiat) < $S$(gas/defi/crypto)

When entropy increases, usually enthalpy decreases. However, they can both increase if the process is endothermic i.e. does not create heat. The entropy of the universe is always increasing. Systems based on maximum entropy can discount prior beliefs and compute markets feedback accurately but there is an associated cost for that. [36]

### 3.6 System optimization by learning agent

Selective machine learning or "gating" consists of introducing an intelligent learning agent for processing incoming signals such as a user's performance score and timing; for computing a selection at the gate; for computing the decisions made after the gate by selected individuals or teams; and for monitoring the retro-signals also known as the effect of the decisions of the users on the ecosystem. [37]

The agent is getting its knowledge from a **double-loop mechanism** between the networks and the gate on one side, and the ecosystem on the other side. The agent is designed for a partially observable environment, stochastic (random in nature), semi-dynamic (the environment itself does not change or very slowly but the agent performance does), continuous (unlimited perceptions), multi-agent operating in the quasi-known environment.

This type of agent has the advantage to start operating in unknown or quasi-known environments and to become more competent than its initial knowledge alone might allow. The most important distinction is made between the learning (element) by short term gate looping which is responsible for making improvements at the gate, and the performance (element or retrograde signaling) learning by long term looping which is responsible for selecting external action's effects and sending a retrograde signal to the gate to **improve selection parameters**.

### 3.7 Social graph of the Nation, a galaxy of many nested graphs

A graph neural network (GNN) is a class of neural network for processing data best represented by graph data structures. Based on the assumption that transactions between users reflect social interactions of a social network and that the density and frequency of transaction are a measurable activity of these users, we are proposing a model of GNN to optimize interactions and increase transactions volume and monitoring between Ydentity ID users in a nested graph. This GNN will be used to predict the **density of transactions** (gas fees paid for consumed transactions) and their **frequency** between the nodes composing a particular nested graph wherein the node features are made from the user-profile which is a Non-Fungible-Token ID number and scores (Ydentity token ID) stored at the user's wallet address.

#### 3.7.1 Problem formulation of nested GNN

Consider a **social graph** deployment comprising N user-profiles. We denote by $\mathbb{F}_t = \{F_{t-T+1}, \ldots, F_t\} \in R^{T \times N}$ a sequence of density of transaction measurements (fees) over T timestamps up to the current time t, where $F_t$ is the transactions snapshot across all user profiles, observed over an interval $[t - \Delta, t]$, i.e., $F_t = \{f_1^t, \ldots, f_N^t\}$, where $f_i^t$ is the density of transactions (fees) at the i-th profile, and $\Delta$ is the temporal granularity of transaction observations configurable by a graph administrator. The social graph is represented as a directed, weighted and dynamic graph. At the t-th time step, we define a graph $G_t = (v_t, A_t)$, where $v_t \in \mathbb{R}^{N \times C}$ is the graph signal and C is the number of features (i.e., density information, user's profile information), and $A_t \in \mathbb{R}^{N \times N}$ the adjacency matrix where an element $a_{i,j}^t$ represents the frequency of transactions between user-profiles i and j observed at that time. Our objective is to predict the most likely density of transactions in the next H time steps, given the past T observations, i.e.,

$$\hat{F}_{t+1}, \ldots, \hat{F}_{t+H} = \arg\max_{v_{t+1},\ldots,v_{t+H}} \log P(F_{t+1}, \ldots, F_{t+H} | G_{t-T+1}, \ldots, G_t).$$

3.7.2 YDentityNet

Inspired from SGDNet [38], we propose **YDentityNet**, a deep neural network that solves the density of transactions forecasting problem posed in Section 3.7.1. YDentityNet captures spatio-temporal correlations among density information at different nodes and dynamic adjacency matrices modelled from frequency information. SGDNet (shown in Fig. 1) consists of a feature extraction block followed by several spatiotemporal (ST) blocks. Each block comprises temporal layers that handle graph signals and adjacency matrices, and spatial layers for dynamic graph convolution. [38] We spotted the similarity between the mobile traffic forecasting problem that SGDNet aims to solve and our nested GNN problem, so we adopted SGDNet as a solution in our case. In what follows we explain these different modules in the model in more detail.

a) Spatiotemporal feature extraction: The first block generates feature maps for the next module by capturing spatiotemporal correlations from graph signals $V \in \mathbb{R}^{T \times N \times C}$ and the dynamic adjacency matrix $A \in \mathbb{R}^{T \times N \times N}$. The feature dimension of $A$ is first reduced by a Convolution Neural Network (CNN) layer before concatenating with $V$, so that features pertaining to $A$ do not dominate in the concatenated matrix. Then the outcome is passed through another CNN layer to extract feature maps for the subsequent spatiotemporal block.

b) Gated TCN: Each ST block encompasses two gated temporal convolution networks (TCNs) and a Dynamic Graph Convolution Network (DGCN). We adopt gated 1-D dilated causal convolution [38] as the Temporal convolution layer to capture complex temporal dependencies. Dilated causal convolution works by sliding over inputs and skipping elements with a periodically increasing step, and it is able to handle long-term sequences in a non-recursive manner. We stacked several dilated causal convolution layers together. Given an input $x \in R^T$ and filter $g \in R^K$, the dilated causal convolution operation step is represented as:

$$x * g(t) = \sum_{s=0}^{K-1} g(s) x(t - d \times s),$$

where d is the dilation factor determining the length of the skipping step. Then we leverage a gating mechanism to control the information flow through layers, [38] as follows:

$$\mathbb{H}(x) = z(x * g_1(t) + b) \odot \sigma(x * g_2(t) + c),$$

where b and c are model weights, $\odot$ is the element-wise multiplication, $z$ ($\cdot$) is an activation function, and $\sigma$ ($\cdot$) is the sigmoid function which controls the information passed to the next layer. We apply Gated TCNs on both inputs $V$ and $A$, to learn their temporal dependencies while reducing the temporal dimension of the propagated output.

c) Dynamic Graph Convolution Network: to obtain accurate predictions both short- and long-term, we combine spectral graph convolution and DCRNN into a dynamic graph convolution network. Applying the spectral graph convolution of $x \in \mathbb{R}^{T \times N \times C}$ and the adjacency matrix $A \in \mathbb{R}^{T \times N \times N}$ along the time dimension is not effective, because these $T$ outcome snapshots share one weight and thus lose temporal correlations. To circumvent this issue, we adopt EvolveGCN, [38] where we assign a weight to each snapshot, and these weights are temporally related by a Gated recurrent unit (GRU). Mathematically, for every snapshot xt and its corresponding adjacency matrix $A_t$,

$$x_t = x_t * \mathcal{G} w_t = \sigma(\tilde{A}_t x_t w_t),$$

$$w_t = \text{GRU}(w_{t-1}),$$

where $\tilde{A}_t = I_n + D^{-\frac{1}{2}} A_t D^{-\frac{1}{2}}$ and $w_t$ is the weight of t-th snapshot. Each GCN operation has a weight, which is generated from the weight in the last snapshot. Compared to the Long Short Term Memory (LSTM), GRU has fewer gates and therefore is faster to train and uses less memory. We initialize $w_1$ at the beginning. For the following recurrent steps, we use the last output as both a hidden state and the input to the GRU. Finally, we concatenate the output from every snapshot as the final output of EvolveGCN. We denote the EvolveGCN operator as $*\mathcal{E}$. We express the DGCN operation in matrix form:

$$\mathbb{Y}(x, A) = x * \mathcal{E} w^e + x * \mathcal{D} w^d$$

$$= \|_{t=1}^{T} \tilde{A}_t x w_t^e + \sum_{k=0}^{K}(P_f^k x w_{k,1}^d + P_b^k x w_{k,2}^d),$$

where $\|$ denotes concatenation, $\tilde{A}_t = I_n + D^{-\frac{1}{2}} A D^{-\frac{1}{2}}$, $w^e \in \mathbb{R}^{T \times C \times C'}$, $w^d \in \mathbb{R}^{C \times C'}$. C′ is the dimension of the hidden states. Finally, the formula of the $l$-th ST block given the input graph signal $v^l \in \mathbb{R}^{T \times N \times C'}$ and input adjacency matrix $A^l \in \mathbb{R}^{T \times N \times N}$, is given by:

$$v^{l+1} = \mathbb{Y}\big(\mathbb{H}(v^l), \mathbb{H}(A^l)\big); A^{l+1} = \mathbb{H}(A^l).$$

d) Training Loss: the purpose of training is to minimize the Mean Square Error (MSE) of every snapshot, i.e.,

$$\mathbb{L}(\hat{F}_{t+1}, \ldots \hat{F}_{t+H}) = \frac{1}{H \times N} \sum_{i=1}^{H} \sum_{n=1}^{N} (\hat{f}_n^i - f_n^i)^2,$$

where H is the number of prediction steps and N is the number of user-profiles in the deployment.

**Figure 1. SDGNet** [38]

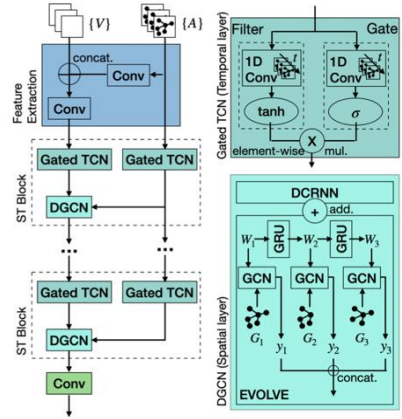

**Nested GNN** will be initially used to optimize the transaction volume and to monitor inside a small graph and will be later generalized to anticipate onboarding rules of new users in the DAN.

Message Passing Neural Networks (MPNN) are models proposed to optimize GNNs for use on larger graphs and apply them to domains such as social graph networks and **online communities**.

## 4. Conclusion

As a reminder of the past, the **Tower of Babel** [39] is a myth referring to misunderstanding between peoples whether it was the wish of a God or the fate of different geographics does not matter, only the lesson remains. We wish we could understand each other, and the web is the network that was built for this particular purpose. In 2000, the internet made headlines as if it "may be a passing fad as millions give up on it". [40] It seems we have passed that test and the majority is using the internet, but questions remain: what do we want to do with it next? And who does it belong to?

The future of value transfer, the future of authorship, the future of work and investment and **the future of the meaning of life** are at stake.

A **Neo-Existentialism** [41] is on its way and about to enter our lives. The time for a reboot has come.

A new form of identity is at the core, but sustainability will come from creating the sense of **connectivity** required in any contract – be it social, political, economic personal or legal – learning from the properties of connectivity observed in the mathematics of thermal dynamics and systems and operated by graphs.

Ydentity enables the creation of an ecosystem or DAN where each Ydentity holder can generate their own "reputation" in the form of YDR reputation tokens that represent the value created by a **positive sum game – of digital life in a decentralized autonomous nation.** Users may choose to liquidate their reputation, but this decision has consequences because YDR tokens can only be earned, and liquidation is irreversible like in real life.

## 5. Acknowledgements

We are grateful to Michael Buxton and Yat Wan Lui for their thoughtful feedback and comments.

Special acknowledgment to PhD candidate Yini Fang for her contribution to the section 3.7 YDentityNet.

All errors and views are our own.

## 6. About the Authors

Dr. Frederic Jumelle, MD, IEEE Senior Member, is a neuroscientist advocating for the revaluation of the approach to personal identity in the digital world for the future of the internet.

Tim Pagett, MAppFin, formerly Deloitte Asia Pacific Financial Services Industry leader, is advocating for the future of value transfer.

Dr. Ryan Lemand, PhD, is an economist advocating for a blockchain economy and a regulated management of crypto assets.

## References

[1] Universal Declaration of Human Rights, United Nations General Assembly resolution 217A (III), Paris, December 10, 1948

[2] International Covenant on Civil and Political Rights, United Nations General Assembly resolution 2200A (XXI), December 16, 1966

[3] Queensland Human Rights Commission, Section 15 of the Human Rights Act 2019

[4] Aadhaar, Unique Identification Authority of India, Government of India, https://uidai.gov.in

[5] Sustainable Development Solutions Network, UN Secretary-General Ban Ki-moon, August 2012. SDG 16.9, "by 2030 provide legal identity for all including free birth registrations."

[6] ISO/IEC 24760-1:2019 "IT Security and Privacy – A framework for identity management – Part 1: Terminology and concepts" May 2019, https://www.iso.org

[7] ID2020 is a digital ID initiative funded by Microsoft and Accenture, https://id2020.org

[8] OpenID, a project to provide the "internet identity layer", https://openid.net/foundation

[9] Lightweight Identity (LID), URL-based identity management system developed by NetMesh and followed by OpenID, OAuth and XDI

[10] G. Greenwald, E. MacAskill, L. Poitras, "Edward Snowden: the whistleblower behind the NSA surveillance revelations", The Guardian, June 11, 2013

[11] C. Cadwalladr, E. Graham-Harrison, "Revealed: 50 million Facebook profiles harvested for Cambridge Analytica in major data breach", The Guardian, March 17, 2018

[12] N. Borenstein, M. Kucherawy, " A Reputation Response Set for Email Identifiers" and "An Architecture for Reputation Reporting", Internet Engineering Task Force, November 2013, https://www.rfc-editor.org/info/rfc7070

[13] Ydentity ID decentralized applications are available at https://web3.ydentity.org and https://web3.rsk.ydentity.info

[14] Association Ydentity.org, Registre du Commerce du Canton de Genève, Switzerland, https://ge.ch/hrcintapp/externalCompanyReport.action?companyOfsUid=CHE-448.613.114&lang=FR

[15] Ydentity Reputation Tokens, symbol YDR, are ERC-20 tokens deployed on contract: https://etherscan.io/address/0x32779a598c7820ad5f13031c0d91e66dceda31bd

[16] Decentralized Identity Foundation (DIF) is a project in favor of a decentralized identity ecosystem and funded by an international consortium of companies and organizations

[17] Self-Sovereign Identities (SSI) are identities controlled by individuals such as the Decentralized Identifiers of W3C

[18] M. S. Ferdous, F. Chowdhury and M. O. Alassafi, "In Search of Self-Sovereign Identity Leveraging Blockchain Technology," in IEEE Access, vol. 7, pp. 103059-103079, 2019, doi: 10.1109/ACCESS.2019.2931173

[19] F. A. Jumelle, Y. W. Lui, "Ydentity (Ydentity Cognitive Theory-based Survey EN/SC)", China work copyright certificate No. 2020-A-01215857

[20] F. A. Jumelle, Y. W. Lui, "Ydentity v1.0 (Ydentity Cognitive Theory-based Survey Design Software)", China software copyright certificate No. 2021SR0038528

[21] F. A. Jumelle, Y. W. Lui, "Method and System for Neuropsychological Performance Test", International Patent (WIPO) published on January 14, 2021, No. WO2021003681A1

[22] F. A. Jumelle, "Method and System for Neuropsychological Performance Test", US Patent application No.17/625,568 of January 7, 2022, as USA national entry of PCT patent No. WO2021003681

[23] F. A. Jumelle, Y. W. Lui, "Method and System for Neuropsychological Performance Test", Hong Kong SAR standard patent certificate No. HK 40052998 A (2021) of January 28, 2022, as published in China on June 18, 2021

[24] F. A. Jumelle, Y. W. Lui, "Method and System for Neuropsychological Performance Test", China Patent application No. 201980072840.X of June 18, 2021, as China national entry of the PCT patent No. WO2021003681Hong Kong Short-Patent no. HK30003521 A was published on 19[th] May 2020

[25] Y. Zhou and B. E. Shi, "Photorealistic Facial Expression Synthesis by the Conditional Difference Adversarial Autoencoder," International Conference on Affective Computing and Intelligent Interaction, San Antonio, TX, USA, Oct. 2017

[26] D. Deng, Z. Chen, B. E. Shi, "Multitask Emotion Recognition with Incomplete Labels," IEEE International Conference on Automatic Face and Gesture Recognition, in Buenos Aires, Argentina, Nov. 2020. (Winner of the Valence-Arousal and the Action Unit Challenges in the FG-2020 Workshop "Affect Recognition in-the-wild: Uni/Multi-Modal Analysis & VA-AU-Expression Challenges)

[27] Y. Zhou, J. Pi and B. E. Shi, "Pose-independent Facial Action Unit Intensity Regression Based on Multi-task Deep Transfer Learning," IEEE International Conference on Automatic Face & Gesture Recognition, Washington, DC, USA, May 2017 (Winner of 2017 Facial Expression

Recognition and Analysis (FERA) AU Intensity Estimation Challenge)

[28] D. Deng, L. Wu, B. E. Shi, "Iterative Distillation for Better Uncertainty Estimates in Multitask Emotion Recognition," IEEE Xplore, November 24, 2021

[29] O. Gokalp and E. Tasci, "Weighted Voting Based Ensemble Classification with Hyper-parameter Optimization," 2019 Innovations in Intelligent Systems and Applications Conference (ASYU), 2019, pp. 1-4, doi: 10.1109/ASYU48272.2019.8946373

[30] Scharfenaker, E. (2022), "Statistical equilibrium methods in analytical political economy," Journal of Economic Surveys, 36: 276-309. https://doi.org/10.1111/joes.12403

[31] Richard H. Thaler [2017 Nobel Memorial Prize in Economic Sciences], Cass R. Sunstein, "Nudge: Improving Decisions About Health, Wealth, and Happiness", Penguin Books, February 24, 2009 (first published April 8, 2008)

[32] Gavin Wood (2015), "Proof of Authority (PoA) Private Chains", GitHub

[33] Anand Banerjee, Victor M. Yakovenko. Universal patterns of inequality. New Journal of Physics. 2010; 12 (7):1.

[34] Ao, P., "Boltzmann Gibbs distribution of fortune and broken time reversible symmetry in Econodynamics", Communications in Nonlinear Science and Numerical Simulations, vol. 12, no. 5, pp. 619–626, 2007. doi: 10.1016/j.cnsns.2005.07.004.

[35] Victor M. Yakovenko, Jr J. Barkley Rosser. Colloquium: Statistical mechanics of money, wealth, and income. Reviews of Modern Physics. 2009; 81 (4):1703-1725

[36] Benjamin Patrick Evans, Mikhail Prokopenko, A Maximum Entropy Model of Bounded Rational Decision-Making with Prior Beliefs and Market Feedback, Entropy, 10.3390/e23060669, 23, 6, (669), (2021).

[37] S. Sadeghi, M. Amiri, and F. M. Mooseloo, "Artificial Intelligence and Its Application in Optimization under Uncertainty", in Data Mining - Concepts and Applications. London, United Kingdom: IntechOpen, 2021 [Online]. Available: https://www.intechopen.com/chapters/78237 doi: 10.5772/intechopen.98628

[38] Y. Fang, S. Ergüt and P. Patras, "SDGNet: A Handover-Aware Spatiotemporal Graph Neural Network for Mobile Traffic Forecasting," in IEEE Communications Letters, vol. 26, no. 3, pp. 582-586, March 2022, doi: 10.1109/LCOMM.2022.3141238.

[39] The "Tower of Babel" narrative, Genesis 11:1-9

[40] Daily Mail, "Internet "may be just a passing fad as millions give up on it', December 5, 2000

[41] Markus Gabriel, "Neo-Existentialism", John Wiley & Sons, November 26, 2018